\documentclass[prd,aps,showpacs,nofootinbib,preprint,eqsecnum]{revtex4}

\usepackage{epsf}
\usepackage{graphics}
\usepackage{amssymb}
\usepackage{amsmath}

\newcommand{\bear}{\begin{array}}  \newcommand{\eear}{\end{array}}
\newcommand{\bea}{\begin{eqnarray}}  \newcommand{\eea}{\end{eqnarray}}
\newcommand{\beq}{\begin{equation}}  \newcommand{\eeq}{\end{equation}}
\newcommand{\bef}{\begin{figure}}  \newcommand{\eef}{\end{figure}}
\newcommand{\bec}{\begin{center}}  \newcommand{\eec}{\end{center}}

\newcommand{\Eqn}[1]{&\hspace{-0.2em}#1\hspace{-0.2em}&}

\def\Vec#1{\mbox{\boldmath $#1$}}
\def\Vecs#1{\mbox{\boldmath\tiny $#1$}}

\begin{document}

\title{Large-scale magnetic fields from inflation due to 
Chern-Simons-like effective interaction}

\author{K.~Bamba, C.~Q.~Geng and S.~H.~Ho
}  
\affiliation{Department of Physics, National Tsing Hua University,
Hsinchu, Taiwan 300}


\begin{abstract}
We discuss the generation of large-scale magnetic fields 
due to the breaking of the conformal invariance in the electromagnetic 
field through the $CPT$-even dimension-six Chern-Simons-like effective 
interaction with a fermion current in inflationary cosmology. 
It is shown that the magnetic fields on 1Mpc scale with the field strength of
$\sim 10^{-9}$G at the present time can be generated even for 
the scale of the effective interaction being the Planck scale. 
\end{abstract}


\pacs{98.80.Cq, 98.62.En}

\maketitle

\section{Introduction}

It is observationally known that 
there exist magnetic fields with the field strength 
$10^{-7}$--$10^{-6}$G on 10kpc--1Mpc scale 
in clusters of galaxies as well as 
$\sim 10^{-6}$G on $1-10$kpc scale in galaxies of 
all types and at cosmological distances 
(for reviews, see~\cite{reviews}). 
However, the origin of the cosmic magnetic fields, in particular 
the large-scale magnetic fields in clusters of galaxies, 
is not well understood yet. 
Although the dynamo amplification mechanism~\cite{EParker} 
amplifies very weak seed magnetic fields up to $\sim 10^{-6}$G, its 
effectiveness 
in galaxies at high redshifts and clusters of galaxies is still unsatisfactory.
Furthermore,  
it is difficult for astrophysical processes~\cite{Biermann1, PI}, 
cosmological phase transitions~\cite{PT} and primordial density perturbations 
before or at the epoch of recombination~\cite{DP} to 
generate the magnetic fields on megaparsec scales with  sufficient field 
strengths to account for the magnetic fields observed in galaxies and 
clusters of galaxies without dynamo amplification mechanism.

The most natural origin of the large-scale magnetic fields is from
electromagnetic quantum fluctuations 
existed at the inflationary 
stage~\cite{Turner:1987bw}. This is because inflation naturally produces 
effects on very large scales, larger than the Hubble horizon, starting from 
microphysical processes operating on a causally connected volume. 
In the Friedmann-Robertson-Walker (FRW) spacetime, the metric is conformally 
flat, while the ordinary Maxwell theory is conformally invariant. 
It is clear that the conformal invariance must have been broken at the 
inflationary stage in order that electromagnetic quantum fluctuations can be 
induced during inflation~\cite{Parker:1968mv}. 
Note that this does not apply when the FRW background has nonzero spatial 
curvature~\cite{NZC}.
Various breaking mechanisms of the conformal invariance in the 
electromagnetic field, 
such as non-minimal gravitational coupling~\cite{Turner:1987bw}, 
coupling to a scalar field~\cite{Ratra:1991bn, DE, C-S, pseudoscalar, 
charged scalar, ScalarED, ScalarED-2, Enqvist:2004yy, DBI}, 
conformal anomaly induced by quantum effects~\cite{Dolgov:1993vg}, 
spontaneous breaking of the Lorentz invariance~\cite{Bertolami:1998dn}, 
noncommutative field theories~\cite{NC}, 
a preferred minimal length~\cite{Ashoorioon:2004rs}, 
and cosmic defects~\cite{Hollenstein:2007kg}, 
have been proposed.

Recently, an $CPT$-even dimension-six Chern-Simons-like
effective interaction between a fermion current and the electromagnetic field
in inflationary cosmology
was proposed to induce the cosmological birefringence~\cite{Geng:2007va, GHN2} 
as well as  baryon number asymmetry~\cite{BGH}.
The electromagnetic quantum fluctuations can be generated during inflation 
as the interaction breaks 
not only the Lorentz invariance but also the conformal invariance 
of the electromagnetic field. 
In the present paper, we study the generation of large-scale magnetic 
fields due to this effective interaction.

This paper is organized as follows.  
In Sec.\ II we describe our model and derive equations of motion for 
the $U(1)$ gauge field. 
In Sec.\ III we consider the evolution of the $U(1)$ gauge field 
and estimate the present strength of the large-scale magnetic fields. 
Finally, Sec.\ IV is devoted to a conclusion. 
We use units 
of $k_\mathrm{B} = c = \hbar = 1$ and adopt 
Heaviside-Lorentz units of electromagnetism.

\section{MODEL}

The action for the Maxwell theory with the $CPT$-even 
dimension-six Chern-Simons-like effective interaction between 
a fermion current ($j_\mu$) and the electromagnetic field 
($A_\mu$)~\cite{Geng:2007va, GHN2}
is given by
\begin{eqnarray}
S 
\Eqn{=}
\int d^{4}x \sqrt{-g} 
\left[ \hspace{1mm}
{\mathcal{L}}_{\mathrm{M}}   
+
{\mathcal{L}}_{\mathrm{CS}}  
\hspace{1mm} \right]\,, 
\label{eq:1}  
\nonumber\\
{\mathcal{L}}_{\mathrm{M}}   
\Eqn{=} -\frac{1}{4} F_{\mu\nu} F^{\mu\nu}\,\ {\mathrm{and}}\
{\mathcal{L}}_{\mathrm{CS}}  
=
-\frac{\beta}{M^2}j_{\mu}A_{\nu}\tilde{F}^{\mu\nu}\,,
\end{eqnarray}
where $g$ is the determinant of the metric tensor $g_{\mu\nu}$, 
$F_{\mu\nu} = \partial_{\mu}A_{\nu}-\partial_{\nu}A_{\mu}$ is 
the electromagnetic field strength tensor, 
$\tilde{F}^{\mu\nu} = \left[ 1/\left(2\sqrt{-g}\right) \right] 
\epsilon^{\mu\nu\rho\sigma} F_{\rho\sigma}$
is the dual of $F_{\mu\nu}$
with $\epsilon^{\mu\nu\rho\sigma}$ being the Levi-Civita tensor 
normalized by  $\epsilon^{0123}=+1$, $\beta$ is a dimensionless 
coupling parameter, and 
$M=\Lambda/4\pi$ with $\Lambda$ being the scale of the effective interaction. 
The Chern-Simons-like effective interaction might originate from the low 
energy effective theory in superstring theory~\cite{Geng:2007va}. 

 From the action in Eq.~(\ref{eq:1}), the equation of motion for 
the electromagnetic field can be derived as 
\begin{eqnarray}
&&
\frac{1}{\sqrt{-g}}{\partial}_{\mu} 
\left( \sqrt{-g} F^{\mu\nu} + \frac{\beta}{M^2}j_{\rho}A_{\sigma}
\epsilon^{\rho\sigma\mu\nu} \right) 
- \frac{\beta}{M^2} j_{\mu}\tilde{F}^{\mu\nu} = 0\,. 
\label{eq:4} 
\end{eqnarray} 
We take the flat 
FRW space-time with the metric, 
$
{ds}^2 = g_{\mu\nu}dx^{\mu}dx^{\nu} 
= -{dt}^2 + a^2(t)d{\Vec{x}}^2, 
$
where $a(t)$ is the scale factor. 
The fermion current $j_{\mu}$ to a comoving observer has the 
following form~\cite{Geng:2007va, GHN2}:
\begin{eqnarray}
\emph{j}_{\mu} = \bar{\psi}\gamma_{\mu}\psi =
\left( n, \Vec{0} \right)\,, \
n \Eqn{\equiv} n_{\psi}-n_{\bar{\psi}}\,, 
\label{eq:6}
\end{eqnarray}
where $n_{\psi}$ and $n_{\bar{\psi}}$ are the number densities of 
the fermion $\psi$ and antifermion $\bar{\psi}$, respectively. 
In the FRW background, for 
the Coulomb gauge of
$A_0(t,\Vec{x}) = 0$ and the case of ${\partial}_j A^j (t,\Vec{x}) =0$, 
Eq.~(\ref{eq:4}) becomes
\begin{eqnarray}
&& 
\ddot{A_i}(t,\Vec{x}) 
+ H \dot{A_i}(t,\Vec{x}) 
- \frac{1}{a^2}{\partial}_j {\partial}_j A_i(t,\Vec{x}) 
+ 2 \frac{\beta}{M^2} n a^{-1}
\epsilon_{ijk} {\partial}_j A_k(t,\Vec{x}) = 0\,, 
\label{eq:7}
\end{eqnarray}
where a dot denotes a time derivative, $H=\dot{a}/a$ is 
the Hubble parameter, and $\epsilon_{ijk}$ is the totally antisymmetric 
tensor $(\epsilon_{123} = +1)$.

\section{Large-scale magnetic fields}
\subsection{Evolution of the $U(1)$ gauge field}

We now consider the case in which a slow-roll exponential inflation 
is realized with the scale factor $a(t)$  given by 
$
a(t) = a_1 \exp \left[ H_{\mathrm{inf}}(t-t_1) \right], 
$
where $a_1$ is the scale factor at the time $t_1$ when a 
comoving wavelength $2\pi/k$ of the $U(1)$ gauge field 
first crosses outside the horizon during 
inflation, $k/(a_1 H_{\mathrm{inf}}) = 1$, and 
$H_{\mathrm{inf}}$ is the Hubble constant at the inflationary stage. 
 From the quantization of the $U(1)$ gauge field $A_{\mu}(t,\Vec{x})$, 
we obtain the expression for $A_i(t,\Vec{x})$ as 
\begin{eqnarray} 
A_i(t,\Vec{x}) \Eqn{=} 
\int \frac{d^3 k}{{(2\pi)}^{3/2}} 
\biggl[ \hat{b}(\Vec{k}) A_i(t,\Vec{k})e^{i \Vecs{k} \cdot \Vecs{x} } 
+{\hat{b}}^{\dagger}(\Vec{k}) 
{A_i^*}(t,\Vec{k})e^{-i \Vecs{k} \cdot \Vecs{x}} \biggr]\,,
\label{eq:8} 
\end{eqnarray}
where $\Vec{k}$ is the comoving wave number, $k$ denotes its amplitude 
$|\Vec{k}|$, and $\hat{b}(\Vec{k})$ and ${\hat{b}}^{\dagger}(\Vec{k})$ 
are the annihilation and creation operators which satisfy 
$
\left[ \hat{b}(\Vec{k}), {\hat{b}}^{\dagger}({\Vec{k}}^{\prime}) \right] = 
{\delta}^3 (\Vec{k}-{\Vec{k}}^{\prime})
 \hspace{1mm}
\mathrm{and}\ \mathrm{others} = 0
$. 
In what follows, we choose the $x^3$ axis to lie along the spatial momentum 
direction \Vec{k} and denote the transverse directions $x^{I}$ with 
$I=1, 2$. 
We use circular polarizations expressed by 
the combination of linear polarizations as 
$A_{\pm}(k,t) \equiv A_1(k,t) \pm i A_2(k,t)$. 
 From Eq.\ (\ref{eq:7}), we find that 
\begin{eqnarray}
&&
\ddot{A}_{\pm}(k,t) 
+ H_{\mathrm{inf}}\dot{A}_{\pm}(k,t) 
+ \left[ 
\left( \frac{k}{a} \right)^2 
\mp 2\frac{\beta}{M^2} n \left( \frac{k}{a} \right) 
\right] A_{\pm}(k,t) = 0\,.  
\label{eq:9}
\end{eqnarray}   

Unfortunately, it is impossible to obtain the analytic solution of 
Eq.\ (\ref{eq:9}) 
for a generic evolution of the fermion number density 
$n$ at the inflationary stage. If $n$ evolves as 
$n \propto a \propto -\eta$, 
where $\eta=\int dt/a(t)$ is conformal time, 
or $n \propto a^{-1}$,
analytic solutions for Eq.\ (\ref{eq:9}) can be derived. 
We will investigate these cases in the next subsection. 
Here we will numerically solve Eq.\ (\ref{eq:9}) at the inflationary stage. 
We assume that the initial amplitudes of $A_+(k,t)$ and $A_-(k,t)$ 
are the same with 
the time $t_1$ as the initial time. 
During inflation ($t_1 \leq t \leq t_{\mathrm{R}}$, where 
$t_{\mathrm{R}}$ is the end of inflation), 
the amplitudes of $A_{\pm}(k,t)$ are expressed as 
$
A_{\pm}(k,t) = C_{\pm}(k,t) A_{\pm}(k,t_1), 
$ 
where $C_{\pm}(k,t)$ are
obtained by numerical calculations with
$C_{\pm}(k,t_1)=1$. 
We note that $C_{\pm}(k,t)$ has the $k$-dependence. Hence, in numerical 
calculations we chose a comoving scale $L= 2\pi/k$. 
By requiring that the vacuum should reduce to the one in Minkowski spacetime 
in the short-wavelength limit, 
we have
$
|A_{\pm}(k,t_1)| = 1/\sqrt{2k}
$ 
and 
$
|\dot{A}_{\pm}(k,t_1)| = H_\mathrm{inf}/\sqrt{2k} 
$.

For $n=\bar{n}a^{-3}$, the numerical results for the evolutions of $C_+(k,t)$ 
and $C_-(k,t)$ are 
shown in Figs.~1 and 2, respectively, 
where $\bar{n}$ is a constant, 
$\xi_n \equiv \left(\bar{n}/\left[ \mathrm{cm}^{-3} \right] \right)^{1/2}/ 
\left( M/\left[ \mathrm{GeV} \right] \right) = 4.53 \times 10^{-44}$, 
$H_{\mathrm{inf}} = 10^{10}$GeV, $\beta=1.0$ and 
a comoving scale $L=2\pi/k = 1\mathrm{Mpc}$.
By using the five year Wilkinson Microwave Anisotropy Probe (WMAP) data 
on the anisotropy of the cosmic microwave background (CMB) 
radiation~\cite{Komatsu:2008hk},
one gets that 
$H_\mathrm{inf} < 6.0 \times 10^{14}$GeV 
from tensor perturbations~\cite{C-H}. 
As shown later, the large-scale 
magnetic fields with the sufficient amplitude at the present time 
can be generated. 
Here, we have used 
$k/a = \exp \left[ - H_{\mathrm{inf}} \left( t-t_1\right)\right] 
H_{\mathrm{inf}}$ and $a_1 = k/H_{\mathrm{inf}}$, 
and 
calculated Eq.\ (\ref{eq:9}) numerically 
at $t=t_1= H_{\mathrm{inf}}^{-1}$ with 
$C_\pm(k,t_1) = 1.0$ and $\dot{C}_\pm(k,t_1) = H_\mathrm{inf}$.

The numerical results  in Figs.~1 and 2 are understood as follows. 
 From Eq.\ (\ref{eq:9})  for $n=\bar{n}a^{-m}$ with $m$ being an integer, the equation of
  $C_{\pm}(k,t)$ is given by 
\begin{eqnarray}
&& \hspace{-6mm}
C_{\pm}^{\prime\prime}\left(k,\tilde{t}\right) + 
C_{\pm}^{\prime}\left(k,\tilde{t}\right) \nonumber \\ 
&& \hspace{10mm}
{}+ \exp\left[ -2\left(\tilde{t}-\tilde{t}_1\right) \right] 
\left\{ 1 \mp J\exp\left[-\left(m-1\right)\left(\tilde{t}-\tilde{t}_1\right) 
\right] \right\} 
C_{\pm}\left(k,\tilde{t}\right) = 0\,, 
\label{eq:I-1}
\end{eqnarray}
where
\begin{eqnarray}
\nonumber
J& =& 2\frac{\beta}{M^2} \frac{1}{H_\mathrm{inf}} \bar{n} a_1^{-m}\,, 
\\
\tilde{t} &\equiv& H_\mathrm{inf} t\,,\  
\tilde{t}_1 = H_\mathrm{inf} t_1,
\label{eq:I-1-2}
\end{eqnarray}
 and the prime denotes the derivative 
with respect to $\tilde{t}$. 
\begin{figure}[tbp]
\begin{center}
   \includegraphics{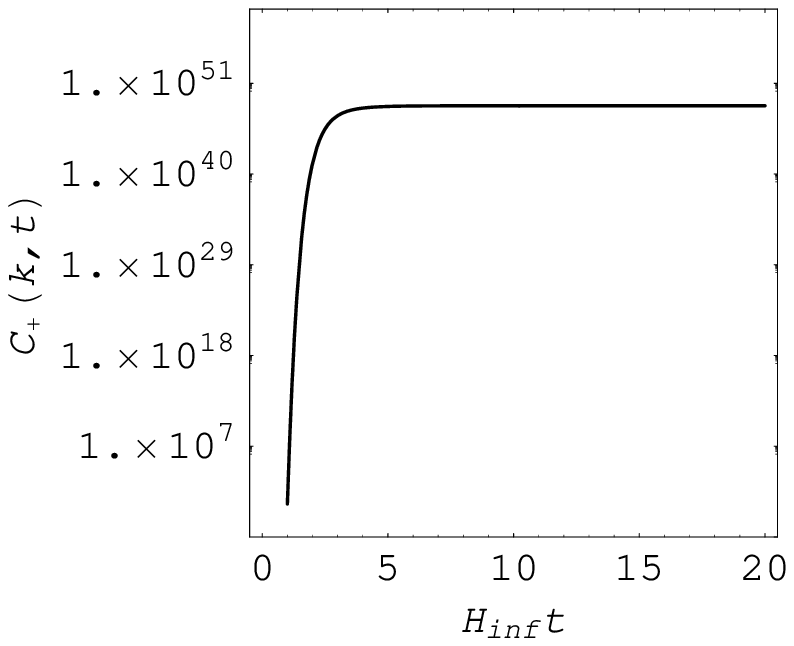}
\caption{
Evolution of $C_+(k,t)$ for 
$n=\bar{n}a^{-3}$, 
$\xi_n = \left(\bar{n}/\left[ \mathrm{cm}^{-3} \right] \right)^{1/2}/ 
\left( M/\left[ \mathrm{GeV} \right] \right) = 4.53 \times 10^{-44}$, 
$H_{\mathrm{inf}} = 10^{10}$GeV, $\beta=1.0$ and 
a comoving scale $L=2\pi/k = 1\mathrm{Mpc}$. 
}
\end{center}
\label{fg:1}
\end{figure}
\begin{figure}[tbp]
\begin{center}
   \includegraphics{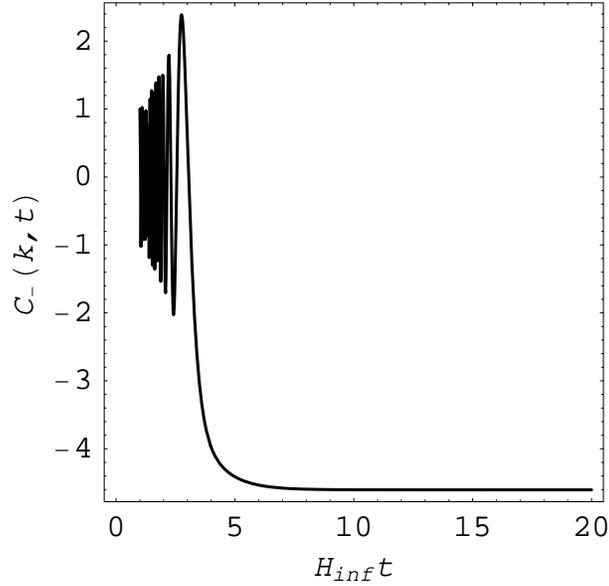}
\caption{
Evolution of $C_-(k,t)$ for 
$n=\bar{n}a^{-3}$. Legend is the same as Fig.~1. \vspace{-8mm}
}
\end{center}
\label{fg:2}
\end{figure}
We now consider the case of $m=3$ shown in Figs.~1 and 2. 
At the early stage of inflation, the second term in the braces $\{ \, \}$ 
on the left-handed side of Eq.\ (\ref{eq:I-1}) is much larger than 
the first one. 
This means that the conformal invariance of the electromagnetic field is 
broken, and hence the amplification of $C_{\pm}\left(k,\tilde{t}\right)$ 
can be realized. 
However, as 
$\tilde{t}$ becomes much larger 
than $\tilde{t}_1$, the second term becomes much smaller than the first one 
due to the existence of the exponential suppression term, 
$\exp\left[-2\left(\tilde{t}-\tilde{t}_1\right) \right]$. 
After about ten Hubble expansion times, 
the second term become negligible comparing with the first one, 
and the above equation becomes almost equal to the 
equation of the ordinary Maxwell theory, which can be  rewritten to 
the following form by replacing the independent variable $t$ with 
the conformal time $\eta$: 
\begin{eqnarray}
\frac{d^2 A_{\pm}(k,\eta)}{d \eta^2} + 
k^2 A_{\pm}(k,\eta) = 0\,.
\label{eq:I-2}
\end{eqnarray}
The solution of Eq.\ (\ref{eq:I-2}) is given by 
$A_{\pm} (k,\eta) = 1/\sqrt{2k} e^{-ik\eta}$. 
Thus, the absolute value of the amplitude is constant as 
$|A_{\pm} (k,\eta)| = 1/\sqrt{2k}$. 
Consequently, 
the solutions of $C_{\pm}(k,t)$ become 
asymptotically constant and  the behaviors of $C_{\pm}(k,t)$ 
 in Figs.~1 and 2 are reasonable. 

\subsection{Strength of the large-scale magnetic fields}

Next, we estimate the present strength of the large-scale magnetic fields. 
The proper magnetic fields are given by~\cite{Ratra:1991bn}
\begin{eqnarray} 
{B}_i^{\mathrm{proper}}(t,\Vec{x})
= a^{-1}{B}_i(t,\Vec{x}) = 
a^{-2}{\epsilon}_{ijk}{\partial}_j A_k(t,\Vec{x})\,,
\label{eq:10}    
\end{eqnarray} 
where ${B}_i(t,\Vec{x})$ is the comoving magnetic field. 
 From Eq.\ (\ref{eq:10}), the energy density
 in Fourier space is 
\begin{eqnarray}
{\rho}_{B}(k,t) \Eqn{=} 
\frac{1}{2}
\left[  
\left| B_+^{\mathrm{proper}}(k,t)
\right|^2 
+
\left| B_-^{\mathrm{proper}}(k,t)
\right|^2 \right]\,, 
\label{eq:11} \\
\left| {B}_\pm^{\mathrm{proper}}(k,t) \right|^2 
\Eqn{=} 
\frac{1}{a^2} \left( \frac{k}{a} \right)^2 |A_\pm(k,t)|^2\,, 
\label{eq:12} 
\end{eqnarray} 
where $B_\pm^{\mathrm{proper}}(k,t) \equiv 
B_1^{\mathrm{proper}}(k,t) \pm i B_2^{\mathrm{proper}}(k,t)$. 
Multiplying ${\rho}_{B}(k,t)$ by the phase-space density of 
$4\pi k^3/(2\pi)^3$, 
we get the energy density of the proper magnetic field as 
\begin{eqnarray}
\hspace{-2mm}
{\rho}_{B}(L,t) = \frac{k^3}{4{\pi}^2}
\left[  
\left| {B}_+^{\mathrm{proper}}(k,t) \right|^2 
+ \left| {B}_-^{\mathrm{proper}}(k,t)
\right|^2 \right]\,, 
\label{eq:13}
\end{eqnarray} 
in the position space on $L$. 
Using Eqs.\ (\ref{eq:11})--(\ref{eq:13}), we find 
\begin{eqnarray} 
{\rho}_{B}(L,t) \Eqn{=} \frac{1}{8\pi^2} \left( \frac{k}{a} \right)^4 
\mathcal{I}(k,t)\,,
\nonumber\\
\mathcal{I}(k,t) \Eqn{\equiv} |C_+(k,t)|^2 + |C_-(k,t)|^2\,, 
\label{eq:15} 
\end{eqnarray}
where $\mathcal{I}(k,t)$ corresponds to the amplification factor 
at the inflationary stage. 
Here, we concentrate on 
the case in which 
after inflation the universe is reheated immediately at 
$t=t_\mathrm{R}$. The conductivity of the universe ${\sigma}_\mathrm{c}$ 
is negligibly small during inflation because there are few charged particles 
at that time. 
After reheating, a number of charged particles are produced 
so that the conductivity immediately jumps to a large value:\ 
${\sigma}_\mathrm{c} \gg H$. 
For a large enough ${\sigma}_\mathrm{c}$, 
magnetic fields evolve in proportion to $a^{-2}(t)$~\cite{Ratra:1991bn}. 
 From $B(L,t) = \sqrt{2{\rho}_{B}(L,t)}$ and 
Eq.\ (\ref{eq:15}), we find that 
the present strength of the magnetic fields is 
\begin{eqnarray}
B(L,t_0) \Eqn{=} 
8.2 \times 10^{18} \exp\left( -2N \right)
\left( \frac{H_{\mathrm{inf}}}{[\mathrm{GeV}]} \right)^2 
\left( \frac{a_{\mathrm{R}}}{a_0} \right)^2 
\sqrt{\mathcal{I}(k,t_{\mathrm{R}})} \hspace{2mm} [\mathrm{G}]\,,
\label{eq:16} 
\end{eqnarray} 
where $
a_{\mathrm{R}}/a_0 =
\left( g_{\mathrm{R}}/3.91 \right)^{-1/3} 
T_{ \gamma 0} / T_{\mathrm{R}}
$
with $T_{\mathrm{R}}$ being the reheating temperature and
$
T_{ \gamma 0} \left( = 2.73 [\mathrm{K}] \right)
$
 the present temperature of the CMB radiation~\cite{Kolb and Turner}, 
$a_\mathrm{R}$ and $a_0 (= 1)$ are the values of $a$ 
at $t = t_{\mathrm{R}}$ and the present time $t_0$, 
 and $N$ is the number of $e$-folds between the time $t_1$ 
and $t_{\mathrm{R}}$, given by 
$N = 45 + \ln \left( L/\mathrm{[Mpc]} \right) + 
\ln J, 
$
where 
$
J= \left[ 30/({\pi}^2 g_\mathrm{R} ) \right]^{1/12} 
{{\rho}_{\mathrm{R}}}^{1/4}/\left(10^{38/3} \mathrm{[GeV]} \right), 
$ 
$g_{\mathrm{R}} (\approx 100)$ is the total number of degrees of freedom for 
relativistic particles at the reheating epoch, and
$
{\rho}_{\mathrm{R}} = 
\left( {\pi}^2/30 \right) g_\mathrm{R} {T_\mathrm{R}}^4$ 
is the energy density of radiation at the 
reheating stage.

Using Eq.\ (\ref{eq:16}) and 
$H_{\mathrm{inf}}^2 = 
\left(8\pi/3\right){\rho}_{\mathrm{R}}/M_{\mathrm{Pl}}^2$, where 
$M_{\mathrm{Pl}}$ is the Planck mass, 
we find that 
for $n=\bar{n}a^{-3}$, 
$\xi_n = 4.53 \times 10^{-44}$, 
$H_{\mathrm{inf}} = 10^{10}$GeV and $\beta=1.0$, 
$C_+(k,t_{\mathrm{R}}) = 1.9 \times 10^{48}$ and 
$C_-(k,t_{\mathrm{R}}) = -4.6$, and consequently 
the field strength of the generated magnetic fields on 
1Mpc scale at the present time is $B_0 (L=1\mathrm{Mpc}, t_0) = 1.1 
\times 10^{-9}$G. 

We account for the relation of the amplitude of the magnetic fields and 
its scale dependence to the evolution of the fermion number density as 
$n=\bar{n}a^{-m}$. It follows from Eq.\ (\ref{eq:I-1}) that 
when the value of $J$ is large, the   breaking magnitude of the conformal 
invariance of the electromagnetic field becomes large and hence 
the strong magnetic fields can be generated. From Eq.\ (\ref{eq:I-1-2}) with 
$a_1 =k/H_{\mathrm{inf}}$, we find 
$
J = 2\beta \left(\bar{n}/M^2\right) \left(1/H_\mathrm{inf}\right) 
\left( k/ H_\mathrm{inf} \right)^{-m} \propto 
\xi_n^2 H_\mathrm{inf}^{m-1} k^{-m}
$. 
When $m>0$, the spectrum of the resultant magnetic fields is red one, 
and the magnetic fields on some given large scale, $i.e.$, small $k$, become 
stronger as $m$ is larger.

For the strength of the primordial magnetic 
fields, there are constraints from Big Bang Nucleosynthesis (BBN) 
on smaller scales. 
The limit on the present strength of the magnetic fields around the BBN horizon 
size $\sim 9.8 \times 10^{-5} h^{-1}\mathrm{Mpc}$ is less than 
$10^{-6}$G~\cite{BBN}. 
Here, $h$ is the related quantity to the present Hubble parameter as 
$H_0=2.13h \times 10^{-42}$GeV~\cite{Kolb and Turner}. 
Note that throughout this paper, we use $h=0.7$~\cite{Freedman:2000cf}.
In the case
in Figs.~1 and 2, the present strength on the 
BBN horizon scale is $5.9 \times 10^{-50} \mathrm{G}$. 
Clearly, the constraints from BBN are satisfied. 
On the other hand, there exist constraints from the CMB anisotropy 
measurements on larger scales. 
The result of $\sim 10^{-9}$G
in Figs.~1 and 2
on 1Mpc 
scale is consistent with the recent observational upper bound derived by 
using the WMAP 5 years data~\cite{Kahniashvili:2008hx}. 
Similar bounds have been studied in Ref.~\cite{CMB-Limit}. 
According to Ref.~\cite{Barrow:1997mj}, 
the limit on the current strength on scales larger than the present horizon 
is less than $4.8 \times 10^{-9} \mathrm{G}$. 
For $\xi_n = 1.87 \times 10^{-49}$, 
$H_{\mathrm{inf}} = 10^{10}$GeV and $\beta=1.0$, the present strength of the 
magnetic fields on the horizon scale is $2.4 \times 10^{-9}$G, which is 
consistent with the above upper limit. In this case, the field 
strength on 1Mpc is $1.2 \times 10^{-57}$G.

In addition, 
we study the case in which $n$ evolves as 
$n=\bar{n}a$. In this case, we can obtain the following analytic solution of 
Eq.\ (\ref{eq:9}): 
\vspace{-2mm}
\begin{eqnarray}
\hspace{-5.5mm}
A_{\pm}(k,a) \Eqn{=} \sqrt{\frac{\pi}{4k} 
\left( \frac{k}{aH_{\mathrm{inf}}} \right)} 
H_{\nu}^{(1)} \hspace{-1mm} \left( \frac{k}{aH_{\mathrm{inf}}} \right) 
e^{i(2\nu+1)\pi/4}, \\
\label{eq:A.7}
\nu \Eqn{=} \sqrt{\frac{1}{4} \pm 2 \frac{\beta}{M^2} \bar{n} k 
\frac{1}{H_{\mathrm{inf}}^2}}\,, 
\label{eq:A.3} 
\end{eqnarray}
where $H_{\nu}^{(1)}$ is a $\nu$th-order Hankel function of type $1$, 
and we have taken $\nu >0$ and 
chosen the integral constant so that the vacuum reduces to 
the one in Minkowski space-time at the short-wavelength limit. 
Being interested in large-scale magnetic fields, 
we investigate the behavior of this solution in the large-wavelength limit. 
Expanding the Hankel function in Eq.\ (\ref{eq:A.7}) and taking the first 
leading order in $k/(a H_{\mathrm{inf}})$, 
from Eqs.\ (\ref{eq:11})--(\ref{eq:13}) we find that 
the present strength of the large-scale magnetic fields is given by 
\begin{eqnarray}
B(L,t_0) \Eqn{=} 
5.1 \times 10^{19} \frac{2^{\nu-1}}{\pi^{3/2}} \Gamma \left( \nu \right) 
\left( \frac{H_{\mathrm{inf}}}{[\mathrm{GeV}]} \right)^2 
\left( \frac{a_{\mathrm{R}}}{a_0} \right)^2  
\left( \frac{k}{aH_{\mathrm{inf}}} \right)^{5/2-\nu} 
\hspace{2mm} [\mathrm{G}]\,. 
\label{eq:A.9} 
\end{eqnarray} 
If $\nu = 2.57$ and $H_{\mathrm{inf}} = 10^{14}$GeV,
the present strength of the large-scale magnetic 
fields on 1Mpc scale is $B(L=1\mathrm{Mpc}, t_0) = 1.4 \times 10^{-9}$G. 
In this case, ${\xi}_n = 3.27 \times 10^{53}$. 

It is interesting to note that for $n = \bar{n} a^{-1}$, 
we can also obtain an analytic solution of Eq.\ (\ref{eq:9}). 
In this case, Eq.\ (\ref{eq:9}) is rewritten to 
the following form by replacing the independent variable $t$ with 
the conformal time $\eta$: 
$
d^2 A_{\pm}\left(\tilde{k},\eta\right)/\left(d \eta^2\right) + 
\tilde{k}^2 A_{\pm}\left(\tilde{k},\eta\right) = 0,
$
where $\tilde{k}^2 = k^2 \mp 2\left(\beta/M^2\right) \bar{n}k$. 
The solution of this equation is given by 
$A_{\pm} (k,\eta) \propto \exp \left(-i \tilde{k} \eta \right)$. 
This is an oscillating solution and its absolute value is constant. 
Hence $A_{\pm}(k,t)$ cannot be amplified.

We remark that 
if $\psi$ is an unknown particle and $n=\bar{n}a^{-3}$, 
the present number density $\bar{n}$ should be smaller than 
that of the neutrino 
$1.1 \times 10^{2} \mathrm{cm}^{-3}$~\cite{Kolb and Turner}. 
For $\bar{n}/\left[ \mathrm{cm}^{-3} \right] = 3.1 \times 10^{-49}$, 
which satisfies the above constraint, 
it follows from 
$\xi_n = \left(\bar{n}/\left[ \mathrm{cm}^{-3} \right] \right)^{1/2}/ 
\left( M/\left[ \mathrm{GeV} \right] \right) = 4.53 \times 10^{-44}$ 
that $M$ can be the Planck scale, $i.e.$, $M = M_{\mathrm{Pl}} = 1.2 
\times 10^{19}$GeV. 

In this paper, we treat the fermion $\psi$ being relativistic during inflation. 
The energy density of the inflaton is given by 
$\rho_{\mathrm{inf}} = 
\left[3/\left(8\pi\right)\right] H_{\mathrm{inf}}^2 M_{\mathrm{Pl}}^2$. 
The energy density of radiation at the reheating stage is given by 
${\rho}_{\mathrm{R}} = 
\left( {\pi}^2/30 \right) g_\mathrm{R} {T_\mathrm{R}}^4$. 
 Since we consider the instantaneous reheating in our study, 
$\rho_{\mathrm{inf}} = {\rho}_{\mathrm{R}}$. 
The energy density of relativistic fermion is given by 
${\rho}_{\mathrm{fermion}} = (7/8) 
\left( {\pi}^2/30 \right) g_\mathrm{R} 
{T_\mathrm{R}}^4$~\cite{Kolb and Turner}. 
Thus, ${\rho}_{\mathrm{fermion}} = (7/8) \rho_{\mathrm{inf}}$, namely, 
the energy density of the fermion $\psi$ is smaller than that of the 
inflaton. 
This is reasonable because in the standard inflationary cosmology 
the potential energy of the inflaton is mainly responsible for inflation. 
Moreover, in this paper we do not study the production mechanism of the 
fermion $\psi$. We assume that it is produced by some other mechanism. 

In addition, we note that 
the present value of the ratio of the Chern-Simons 
interaction term to the Maxwell one is given by 
$|{\mathcal{L}}_{\mathrm{CS}}/{\mathcal{L}}_{\mathrm{M}}| \approx 
10 \beta {\xi}_n^2$, where 
we have used $|\partial_{\mu}A_{\nu}/A_{\nu}| 
\approx H_0$. 
Hence, for $n=\bar{n}a^{-3}$, 
$\xi_n = 4.53 \times 10^{-44}$ 
and $\beta=1.0$, the above ratio is 
much smaller than unity. Thus, the relative contribution of the 
Chern-Simons interaction term at the present time is very small. 

\section{Summary}

In summary, we have studied the generation of the large-scale magnetic 
fields due to the breaking of the conformal invariance in the electromagnetic 
field through the $CPT$-even dimension-six Chern-Simons-like effective 
interaction with a fermion current in inflationary cosmology. 
We have found that the magnetic fields on 1Mpc scale 
with the present amplitude of $\sim 10^{-9}$G can be generated. 
This strength is enough to explain the magnetic fields observed in galaxies and clusters of galaxies through only adiabatic compression without requiring 
any dynamo amplification~\cite{Turner:1987bw}. If the number density of the 
fermion $\psi$ interacting with the electromagnetic field evolves in 
proportion to $a^{-3}(t)$ during inflation, the scale of the effective 
interaction can be the Planck scale.

\begin{acknowledgments}
This work is supported in part by 
the National Science Council of R.O.C. under 
Grant \#: 
NSC-95-2112-M-007-059-MY3 and 
National Tsing Hua University under Grant \#: 
97N2309F1.
\end{acknowledgments}

\textit{Note added.}---
After this work was completed, we became aware of 
a related work~\cite{Campanelli:2008tt} considering the generation 
of the primordial magnetic fields during inflation in a Lorentz violating 
theory of Electrodynamics due to a Chern-Simons term coupling the 
$U(1)$ gauge field to an external four-vector, 
proposed by Carroll, Field and Jackiw~\cite{Carroll:1989vb}. 
In the scenario of Ref.~\cite{Campanelli:2008tt}, during inflation 
the induced magnetic fields are peaked on much smaller scales than the Hubble horizon at that time. 
To extend the scales to 
the galactic one at the time of protogalactic collapse, 
the inverse cascade mechanism has to work. 
On the other hand, in our scenario, during inflation the magnetic fields can be generated on much larger scales than the Hubble horizon at that time 
so that the magnetic fields with sufficient strength on the scale of 
cluster of galaxies at the present time are produced without any 
secondary extension mechanism such as the inverse cascade. This is the 
advantaged feature of our scenario.


\end{document}